\documentclass[12pt]{article}

\begin{document}

\title{Some new integrable systems of two-component fifth order equations}

\author{ {\normalsize Daryoush Talati $^{a}$%
,\, \thanks{%
Talati@eng.ankara.edu.tr,;~Daryoush.Talati@gmail.com}}\,\, {Abdul-Majid Wazwaz $^b$
,\, \thanks{%
Wazwaz@sxu.edu}}
\\
$^a$ {\small Department of Engineering Physics, Ankara University 06100 Tandogan, Ankara, Turkey.}\\
$^b$ {\small Department of Mathematics, Saint Xavier University, Chicago, IL
60655 USA } }

\maketitle

\newtheorem{theo}{Theorem}[section]
\newtheorem{prop}{Proposition}[section]
\newtheorem{con}{Conjecture}[section]
\begin{abstract}
In this work we develop some   fifth-order integrable coupled systems of weight $0$ and $1$ which possess seventh-order symmetry. We establish four new  systems, where in some cases, related recursion operator and bi-Hamiltonian formulations are given.   We also investigate the integrability of the developed systems.

\end{abstract}

\section{Introduction}

   Integrable  systems of equations, that possess sufficiently  large number of conservation laws and give rise to multiple soliton solutions    play a major role in theoretical physics and in propagation of waves.  The work on  integrable  systems of equations is flourishing  because  these systems  have richer phenomena in scientific applications  than the regular systems.

An evolution equation is defined to be integrable in symmetry sense if it admits infinitely many symmetries. Integrable systems are nonlinear differential equations which can be solved analytically.  Exactly solvable models and integrable evolution equations in nonlinear science
 play an essential role in many branches of science and engineering.  The useful findings in integrable  systems of equations have stimulated much research activity.

The study of constructing integrable  systems of equations  by using methods, such as  recursion operator, symmetries,  bi-Hamiltonian,  and others,    is an interesting topic of growing interest and has gained large interest recently.    Magri \cite{Magri (1980)}   studied the connection between conservation laws and symmetries from the geometric point of view,   where he   proved that some systems admitted two distinct but compatible Hamiltonian structures, now known as bi-Hamiltonian system.

In recent years studies on fifth-order systems of two-component nonlinear evolution equations have received considerable attention \cite{kac,tlax,tt11}. Multi-component generalizations of fifth order Kaup-Kupershmidt equation
\begin{eqnarray}
u_t=u_{5x}+10uu_{3x}+25u_{x}u_{xx}+20u^2u_{x},
\end{eqnarray}
Sawada-Kotera equation
\begin{eqnarray}
u_t=u_{5x}+5uu_{3x}+ 5u_{x}u_{xx}+5u^2u_{x},
\end{eqnarray}
and Kupershmidt equation
\vspace{-0.3cm}
\begin{eqnarray}
u_t=u_{5x}+5u_{x}u_{3x}+5u_{xx}^2-5u^2u_{3x}-20uu_{x}u_{xx}-5u_{x}^3+5u^4u_{x}.
\vspace{-1cm}
\end{eqnarray}
have been the subject of systematic integrability study. Among these, only five homogeneous systems of two-component cases have been found \cite{Mikhailov et al. (2007), Mikhailov et al. (2009),Talati (2013)} so far. Here we mention papers pertaining to multi-component generalizations of fifth order systems only. For the other integrable systems and their properties, we refer the readers to the useful  papers \cite{classi1,classi2,classi3,classi4} and the some of the  references therein. 

So far the only known integrable systems of fifth order two-component equations are as follows

\begin{eqnarray}
\left(
\begin{array}{l}
u\\ \\
v\\ \\
\end{array}\right)_{t}=
\left(
\begin{array}{cc} 
-\frac{5}{3}u_{5x} -10vv_{3x}+10uu_{3x}+ 25u_{x}u_{xx} - 15v_{x}v_{xx} -12u^2u_{x}\\+6v^2u_{x}+12uvv_{x}-6v^2v_{x}\\
15v_{5x} -10vu_{3x}-30uv_{3x}-35v_{x}u_{xx} +30v_{x}v_{xx} -45u_{x}v_{xx}\\ +6v^2u_{x}-6v^2v_{x} + 12uvu_{x} + 12u^2v_{x} \label{sys1}
\end{array}
\right), \label{a1}
\end{eqnarray}

\begin{eqnarray}
\left(
\begin{array}{l}
u\\ 
v\\ 
\end{array}\right)_{t}=
\left(
\begin{array}{cc} 
u_{5x} + 10uu_{3x} + 25u_{x}u_{xx}+20u^2u_{x}+v^2v_{x}\\
u_{3x}v + u_{xx}v_{x}+8uvu_{x}+4u^2v_{x}\label{sys2}
\end{array}
\right), \label{a2}
\end{eqnarray}

\begin{eqnarray}
\left(
\begin{array}{l}
u\\ 
v\\ 
\end{array}\right)_{t}=
\left(
\begin{array}{cc} 
-\frac{1}{8}u_{5x}-2 uu_{3x}-2u_{x}u_{xx}-\frac{32}{5}u^2u_{x}+v_{x}\\
\frac{9}{8}v_{5x} + 6uv_{3x} + 6u_{x}v_{xx} + 4u_{xx}v_{x}+ \frac{32}{5}u^2v_{x}\label{sys3}
\end{array}
\right),\label{a3}
\end{eqnarray}

\begin{eqnarray}
\left(
\begin{array}{l}
u\\ \\\\
v\\ \\
\end{array}\right)_{t}=
\left(
\begin{array}{cc} 

u_{5}+\frac{5}{2}v_{5}+6u_{3}u+18u_{3}v+12v_{3}u+42v_{3}v+12u_{2}u_{1}+24u_{2}v_{1}+21v_{2}u_{1}\\+42v_{2}v_{1}+\frac{54}{5}u_{1}u^2+\frac{108}{5}u_{1}uv-18u_{1}v^2+\frac{72}{5}v_{1}u^2-\frac{72}{5}v_{1}uv-144v_{1}v^2 
\\\\
\frac{5}{4}u_{5}+\frac{7}{2}v_{5}+3u_{3}u+6v_{3}u-6v_{3}v+\frac{3}{2}u_{2}u_{1}-6u_{2}v_{1}-3v_{2}u_{1}-33v_{2}v_{1}\\+\frac{36}{5}u_{1}v^2-\frac{18}{5}v_{1}u^2-\frac{36}{5}v_{1}uv+\frac{126}{5}v_{1}v^2

\end{array}
\right),\label{a4}
\end{eqnarray}

and

\begin{eqnarray}
\left(
\begin{array}{l}
u\\\\\\\\\\\\\\
v\\\\\\\\\\\\
\end{array}\right)_{t}=
\left(
\begin{array}{cc} 
u_{5x}-30vv_{4x}+5u_{x}u_{3x}-5u^{2}u_{3x}+15v^{2}u_{3x}-75v_{x}v_{3x}\\+60uvv_{3x} +90v^{2}v_{3x}+5u_{xx}^{2}-20uu_{x}u_{xx}+60vv_{x}u_{xx}-45v_{xx}^{2}\\+90vu_{x}v_{xx} +90uv_{x}v_{xx}+540vv_{x}v_{xx}+30u^{2}vv_{xx}-180uv^{2}v_{xx}\\-90v^{3}v_{xx}-5u_{x}^{3}+45u_{x}v_{x}^{2}+60uvu_{x}v_{x}-180v^{2}u_{x}v_{x} +5u^{4}u_{x} \\-90u^{2}v^{2}u_{x} +45v^{4}u_{x}+180v_{x}^{3}+30u^{2}v_{x}^{2} -360uvv_{x}^{2} -270v^{2}v_{x}^{2} \\ -60u^{3}vv_{x}+180uv^{3}v_{x} \\ \\

-9v_{5x}+10vu_{4x}+25v_{x}u_{3x}+20uvu_{3x}+30v^{2}u_{3x}+15u_{x}v_{3x}\\+90v_{x}v_{3x} +15u^{2}v_{3x}+15v^{2}v_{3x}+30u_{xx}v_{xx}+50vu_{x}u_{xx}\\-10u^{2}vu_{xx}+50uv_{x}u_{xx} +60vv_{x}u_{xx}+60uv^{2}u_{xx}+30v^{3}u_{xx}\\+90v_{xx}^{2} +60uu_{x}v_{xx}+60vv_{x}v_{xx} +45u_{x}^{2}v_{x}-20uvu_{x}^{2}\\+60v^{2}u_{x}^{2}-10u^{2}u_{x}v_{x} +90v^{2}u_{x}v_{x} -20u^{3}vu_{x} +120uvu_{x}v_{x}\\+60uv^{3}u_{x}+15v_{x}^{3} -5u^{4}v_{x} +90u^{2}v^{2}v_{x} -45v^{4}v_{x}
\label{new}
\end{array}
\right). \label{a5}
\end{eqnarray}
%%%%%%%%%%%%%%%%%%%%%%%%%%%%%%%%%%%%%%%%%%%%%%%%%%%%%%%%77%%%%%%%%%%%%%%%%%%%%%%%%%%%%%%%%%%%%%%%%%%%%%%%%%%%%%%%%%%%%%%%%%%%%%%%%%%%%
%%%%%%%%%%%%%%%%%%%%%%%%%%%%%%%%%%%%%%%%%%%%%%%%%%%%%%%%77%%%%%%%%%%%%%%%%%%%%%%%%%%%%%%%%%%%%%%%%%%%%%%%%%%%%%%%%%%%%%%%%%%%%%%%%%%%%
%%%%%%%%%%%%%%%%%%%%%%%%%%%%%%%%%%%%%%%%%%%%%%%%%%%%%%%%77%%%%%%%%%%%%%%%%%%%%%%%%%%%%%%%%%%%%%%%%%%%%%%%%%%%%%%%%%%%%%%%%%%%%%%%%%%%%
Bi-Hamiltonian structures and recursion operators  for the aforementioned  systems are discussed  in \cite{tcomplete,Vojcak (20011)} and in some of the  references therein. System (\ref{sys1}) and (\ref{sys2}) admit a reduction $v=0$ to the Kaup-Kupershmidt eqnarray. By setting $v=0,$ system (\ref{sys3}) reduces to the Sawada-Kotera equation. By setting $v=0$ the well known Kupershmidt equation is an obvious reduction of system (\ref{new}).

%%%%%%%%%%%%%%%%%%%%%%%%%%%%%%%%%%%%%%%%%%%%%%%%%%%%%%%%%%%%%%%%%%%%%%%%%%%%%%%%%%%%%%%%%%%%%%%%%%%%%%%%%%%%%%%%%%%%%%%%%%%%%%%%%%%%%%%%%%%%%%%%%%%%%%%%%%%%%%%%%%%%%%%%%%%%%%%%%%%%%%%%%%%%%%%%%%%%%%%%%%%%%%%%%%%%%%%%%%%%%%%%%%%%%%%%%%%%%%%%%%%%%%%%%%%%%%%%%%%%%%%%%%%%%%%%%%%%%%%%%%%%%%%%%%%%%%%%%%%%%%%%%%%%%%%%%%%%%%%%%%%%%%%%%%%%%%%%%%%%%%%%%%%%%%%%%%%%%%%%%%%%%%%%%%%%%%%%%%%%%%%%%%%%%%%%%%%%%%%%%%%%%%%%%%%%%%%%%%%%%%%%%%%%%%%%%%%%%%%%%%%%%%%%%%%%%%%%%%%%%%%%%%%%%%%%%%%%%%%%%%%%%%%%%%%%%%%%%%%%%%%%%%%%%%%%%%%%%%%%%%%%%%%%%%%%%%%%%%%%%%%%%%%%%%%%%%%%%%%%%%%%%%%%%%%%%%%%%%%%%%%%%%%%%%%%%%%%%%%%

$\\ \\\\\\ \\\\\\$

\section{New homogeneous fifth-order integrable Systems}

In the literature, all of classified integrable systems are second and third order generalization of the KdV and Burgers equations, or equations related to the KdV  and Burgers equations. In the case of fifth-order systems, because of the very big number of arbitrary terms that must be considered, the act of classification of such systems is  very complicated. Motivated by some existing  examples of bi-Hamiltonian two-component generalization of fifth-order equations, we considered a narrow class of fifth-order two-component systems with specific Jordan matrix for integrability.

From a practical point of view, we observed that in second and third order integrable systems, when there is a 2-homogeneous integrable equation in a specific Jordan form, then there is certainly at least one ${1,0}$-homogeneous system in that Jordan form. Then using the sense of  2-homogeneous fifth order systems introduced in \cite{Mikhailov et al. (2007), Mikhailov et al. (2009)},  we aim to develop  new integrable ${1,0}$-homogeneous systems in the  same Jordan form. Our analysis  found four new integrable systems, where  some of these systems allow us  to write Magri schemes which contain the new systems proving it  complete integrability. In what follows, we introduce the new fifth order two-component systems with the form

\begin{eqnarray} 
\left(
\begin{array}{l}
u\\ \\\\\\ \\\\\\
v\\ \\\\ \\\\\\
\end{array}\right)_{t}=
\left(
\begin{array}{cc}
4u_5+5v_5+20u_4u_1+10u_4v_1+40u_1v_4+20v_4v_1+20u_3u_2-40u_3u_1^2\\+140u_3u_1v_1+70u_3v_2+80u_3v_1^2+40u_2^2u_1+80u_2^2v_1-80u_2u_1^3\\+360u_2u_1^2v_1+400u_2u_1v_2+600u_2u_1v_1^2+70u_2v_3+260u_2v_2v_1+200u_2v_1^3\\+24u_1^5-240u_1^4v_1-160u_1^3v_2+360u_1^3v_1^2+40u_1^2v_3+720u_1^2v_2v_1\\+1200u_1^2v_1^3+220u_1v_3v_1+370u_1v_2^2+1200u_1v_2v_1^2+600u_1v_1^4+110v_3v_2\\+100v_3v_1^2+200v_2^2v_1+400v_2v_1^3\\\\
10u_5+14v_5-40u_4u_1-20u_4v_1-20u_1v_4+20u_3u_2-40u_3u_1^2\\-40u_3u_1v_1 +100u_3v_2-100u_3v_1^2-200u_2^2u_1-40u_2^2v_1\\+160u_2u_1^3-720u_2u_1^2v_1-560u_2u_1v_2-1200u_2u_1v_1^2+100u_2v_3\\-400u_2v_2v_1-400u_2v_1^3+600u_1^4v_1+80u_1^3v_2+1200u_1^3v_1^2-200u_1^2v_3\\-360u_1^2v_2v_1+360u_1^2v_1^3-380u_1v_3v_1-320u_1v_2^2-600u_1v_2v_1^2\\-240u_1v_1^4-10v_4v_1+140v_3v_2-320v_3v_1^2-370v_2^2v_1-200v_2v_1^3+24v_1^5
\end{array}\right),\label{es1},
\end{eqnarray}

\begin{eqnarray} 
\left(
\begin{array}{l}
u\\ \\\\\\ \\
v\\ \\\\ 
\end{array}\right)_{t}=
\left(
\begin{array}{cc}
4u_{5x}+4v_{4x}v+20u_{3x}u_{x}-20u_{3x}u^2-8u_{3x}v^2+16v_{3x}v_{x}-8v_{3x}uv+20u_{xx}^2\\-80u_{xx}u_{x}u-8u_{xx}v_{x}v-6u_{xx}uv^2+12v_{xx}^2-12v_{xx}u_{x}v-24v_{xx}v_{x}u\\-4v_{xx}u^2v-8v_{xx}v^3-20u_{x}^3-12u_{x}^2v^2-12u_{x}v_{x}^2-8u_{x}v_{x}uv+20u_{x}u^4\\+24u_{x}u^2v^2+u_{x}v^4-4v_{x}^2u^2-12v_{x}^2v^2+8v_{x}u^3v+10v_{x}uv^3\\\\
-2u_{4x}v+4u_{3x}v_{x}+2u_{3x}uv-2v_{3x}v^2-10u_{xx}u_{x}v-4u_{xx}v_{x}u+8u_{xx}u^2v\\+4u_{xx}v^3-2v_{xx}v_{x}v+6v_{xx}uv^2+12u_{x}^2v_{x}+16u_{x}^2uv-16u_{x}v_{x}u^2\\+6u_{x}v_{x}v^2-8u_{x}u^3v+8u_{x}uv^3+4v_{x}^3-6v_{x}^2uv+4v_{x}u^4+5v_{x}v^4\\

\end{array}\right),\label{es2},
\end{eqnarray}

\begin{eqnarray}
\left(
\begin{array}{l}
u\\ \\\\\\\\\\\\\\
v\\ \\\\\\\\\\\\
\end{array}\right)_{t}=
\left(
\begin{array}{cc}
{u}_{5x}+{v}_{5x}+2{u}_{x}{u}_{4x}-2{v}_{x}{u}_{4x}+6{u}_{x}{v}_{4x}-6{v}_{x}{v}_{4x} -16{u}_{xx}{u}_{3x}\\ -4{v}_{xx}{u}_{3x} -54{u}_{x}^{2}{u}_{3x} -20{u}_{x}{v}_{x}{u}_{3x}-6{v}_{x}^{2}{u}_{3x}-4{u}_{xx}{v}_{3x} \\-16{v}_{xx}{v}_{3x}-22{u}_{x}^{2}{v}_{3x} -52{u}_{x}{v}_{x}{v}_{3x} -6{v}_{x}^{2}{v}_{3x} -52{u}_{x}{u}_{xx}^{2} -4{v}_{x}{u}_{xx}^{2}\\-32{u}_{x}{u}_{xx}{v}_{xx}-16{v}_{x}{u}_{xx}{v}_{xx} -12{u}_{x}^{3}{u}_{xx} +4{u}_{x}^{2}{v}_{x}{u}_{xx} -4{u}_{x}{v}_{x}^{2}{u}_{xx}\\ +12{v}_{x}^{3}{u}_{xx}-44{u}_{x}{v}_{xx}^{2}-12{v}_{x}{v}_{xx}^{2} -36{u}_{x}^{3}{v}_{xx} +12{u}_{x}^{2}{v}_{x}{v}_{xx}\\-12{u}_{x}{v}_{x}^{2}{v}_{xx} +36{v}_{x}^{3}{v}_{xx}+72{u}_{x}^{5}+96{u}_{x}^{4}{v}_{x} +176{u}_{x}^{3}{v}_{x}^{2} +96{u}_{x}^{2}{v}_{x}^{3}\\+72{u}_{x}{v}_{x}^{4}
\\\\
{u}_{5x}+{v}_{5x}-6{u}_{x}{u}_{4x}+6{v}_{x}{u}_{4x}-2{u}_{x}{v}_{4x}+2{v}_{x}{v}_{4x}-16{u}_{xx}{u}_{3x} \\-4{v}_{xx}{u}_{3x} -6{u}_{x}^{2}{u}_{3x} -52{u}_{x}{v}_{x}{u}_{3x} -22{v}_{x}^{2}{u}_{3x}-4{u}_{xx}{v}_{3x}\\-16{v}_{xx}{v}_{3x} -6{u}_{x}^{2}{v}_{3x} -20{u}_{x}{v}_{x}{v}_{3x}-54{v}_{x}^{2}{v}_{3x} -12{u}_{x}{u}_{xx}^{2}\\-44{v}_{x}{u}_{xx}^{2} -16{u}_{x}{u}_{xx}{v}_{xx} -32{v}_{x}{u}_{xx}{v}_{xx} +36{u}_{x}^{3}{u}_{xx}\\-12{u}_{x}^{2}{v}_{x}{u}_{xx} +12{u}_{x}{v}_{x}^{2}{u}_{xx} -36{v}_{x}^{3}{u}_{xx} -4{u}_{x}{v}_{xx}^{2}-52{v}_{x}{v}_{xx}^{2} \\+12{u}_{x}^{3}{v}_{xx} -4{u}_{x}^{2}{v}_{x}{v}_{xx} +4{u}_{x}{v}_{x}^{2}{v}_{xx} -12{v}_{x}^{3}{v}_{xx} +72{u}_{x}^{4}{v}_{x}\\+96{u}_{x}^{3}{v}_{x}^{2} +176{u}_{x}^{2}{v}_{x}^{3} +96{u}_{x}{v}_{x}^{4}+72{v}_{x}^{5}
\end{array}\right),\label{s1}
\end{eqnarray}
and 

\begin{eqnarray}
\left(
\begin{array}{l}
u\\ \\\\\\\\\\\\\\
v\\ \\\\\\\\\\\\
\end{array}\right)_{t}=
\left(
\begin{array}{cc}
{u}_{5x}+{v}_{5x}+{u}_{x}{u}_{4x}-{v}_{x}{u}_{4x}+3{u}_{x}{v}_{4x}-3{v}_{x}{v}_{4x}+7{u}_{xx}{u}_{3x}\\+13{v}_{xx}{u}_{3x} -36{u}_{x}^{2}{u}_{3x} -20{u}_{x}{v}_{x}{u}_{3x} -24{v}_{x}^{2}{u}_{3x}+13{u}_{xx}{v}_{3x}\\+7{v}_{xx}{v}_{3x} -28{u}_{x}^{2}{v}_{3x}-28{u}_{x}{v}_{x}{v}_{3x}-24{v}_{x}^{2}{v}_{3x} -28{u}_{x}{u}_{xx}^{2}\\-16{v}_{x}{u}_{xx}^{2}-8{u}_{x}{u}_{xx}{v}_{xx} -64{v}_{x}{u}_{xx}{v}_{xx}-24{u}_{x}^{3}{u}_{xx}+8{u}_{x}^{2}{v}_{x}{u}_{xx}\\ -8{u}_{x}{v}_{x}^{2}{u}_{xx} +24{v}_{x}^{3}{u}_{xx}+4{u}_{x}{v}_{xx}^{2}-48{v}_{x}{v}_{xx}^{2}-72{u}_{x}^{3}{v}_{xx}\\ +24{u}_{x}^{2}{v}_{x}{v}_{xx} -24{u}_{x}{v}_{x}^{2}{v}_{xx} +72{v}_{x}^{3}{v}_{xx} +72{u}_{x}^{5}+96{u}_{x}^{4}{v}_{x}\\+176{u}_{x}^{3}{v}_{x}^{2} +96{u}_{x}^{2}{v}_{x}^{3} +72{u}_{x}{v}_{x}^{4}
\\\\
{u}_{5x}+{v}_{5x}-3{u}_{x}{u}_{4x}+3{v}_{x}{u}_{4x}-{u}_{x}{v}_{4x}+{v}_{x}{v}_{4x}+7{u}_{xx}{u}_{3x}\\+13{v}_{xx}{u}_{3x}-24{u}_{x}^{2}{u}_{3x} -28{u}_{x}{v}_{x}{u}_{3x}-28{v}_{x}^{2}{u}_{3x}+13{u}_{xx}{v}_{3x}\\+7{v}_{xx}{v}_{3x} -24{u}_{x}^{2}{v}_{3x}-20{u}_{x}{v}_{x}{v}_{3x} -36{v}_{x}^{2}{v}_{3x} -48{u}_{x}{u}_{xx}^{2}\\+4{v}_{x}{u}_{xx}^{2} -64{u}_{x}{u}_{xx}{v}_{xx}-8{v}_{x}{u}_{xx}{v}_{xx}+72{u}_{x}^{3}{u}_{xx} -24{u}_{x}^{2}{v}_{x}{u}_{xx}\\+24{u}_{x}{v}_{x}^{2}{u}_{xx} -72{v}_{x}^{3}{u}_{xx}-16{u}_{x}{v}_{xx}^{2}-28{v}_{x}{v}_{xx}^{2}+24{u}_{x}^{3}{v}_{xx} \\-8{u}_{x}^{2}{v}_{x}{v}_{xx} +8{u}_{x}{v}_{x}^{2}{v}_{xx}-24{v}_{x}^{3}{v}_{xx} +72{u}_{x}^{4}{v}_{x}+96{u}_{x}^{3}{v}_{x}^{2} \\+176{u}_{x}^{2}{v}_{x}^{3} +96{u}_{x}{v}_{x}^{4}+72{v}_{x}^{5}
\end{array}\right)\label{s2}.
\end{eqnarray}

%%%%%%%%%%%%%%%%%%%%%%%%%%%%%%%%%%%%%%%%%%%%%%%%%%%%%%%%%%%%%%%%%%%%%%%%%%%%%%%%%%%%%%%%%%%%%%%%%%%%%%%%%%%%%%%%%%%%%%%%%%%%%%%%%%%%%%%%%%%%%%%%%%%%%%

To find  the second  set of systems, we use a classification that we restricted  to the case  $\lambda=0$ homogeneous symmetrically coupled systems. We determine all equations of the form as

\begin{eqnarray}
\left(
\begin{array}{l}
u\\
v\\
\end{array}\right)_{t}=
\left(
\begin{array}{cc}
A[u_x,v_x] \\
A[v_x,u_x]
\end{array}\right).\label{pot}
\end{eqnarray}

 with The class of two-component 0-homogeneous symmetrically coupled systems with undetermined constant coefficients $\gamma_{i}$ have the form

\begin{eqnarray}
\begin{array}{ll}
A=&{\gamma}_{1} u_{5x}+{\gamma}_{2} v_{5x}+{\alpha}_{1}u_{x}u_{4x}+{\alpha}_{2}v_{x}u_{4x}+{\alpha}_{3}u_{x}v_{4x}+{\alpha}_{4}v_{x}v_{4x}+{\alpha}_{5}u_{xx}u_{3x}\\&+{\alpha}_{6}v_{xx}u_{3x}+{\alpha}_{7}u_{x}^2u_{3x}+{\alpha}_{8}u_{x}v_{x}u_{3x}+{\alpha}_{9}v_{x}^2u_{3x}+{\alpha}_{10}u_{xx}v_{3x}+{\alpha}_{11}v_{xx}v_{3x}\\&+{\alpha}_{12}u_{x}^2v_{3x}+{\alpha}_{13}u_{x}v_{x}v_{3x}+{\alpha}_{14}v_{x}^2v_{3x}+{\alpha}_{15}u_{x}u_{xx}^2+{\alpha}_{16}v_{x}u_{xx}^2\\&+{\alpha}_{17}u_{x}v_{xx}u_{xx}+{\alpha}_{18}v_{x}v_{xx}u_{xx}+{\alpha}_{19}u_{x}^3u_{xx}+{\alpha}_{20}u_{x}^2v_{x}u_{xx}+{\alpha}_{21}u_{x}v_{x}^2u_{xx}\\&+{\alpha}_{22}v_{x}^3u_{xx} +{\alpha}_{23}u_{x}v_{xx}^2 +{\alpha}_{24}v_{x}v_{xx}^2+{\alpha}_{25}u_{x}^3v_{xx}+{\alpha}_{26}u_{x}^2v_{x}v_{xx}\\&+{\alpha}_{27}u_{x}v_{x}^2v_{xx}+{\alpha}_{28}v_{x}^3v_{xx} +{\alpha}_{29}u_{x}^5+{\alpha}_{30}u_{x}^4v_{x} +{\alpha}_{31}u_{x}^3v_{x}^2+{\alpha}_{32}u_{x}^2v_{x}^3\\&+{\alpha}_{33}u_{x}v_{x}^4+{\alpha}_{34}v_{x}^5
\end{array}\label{five}
\end{eqnarray}
possessing an admissible generator of form (\ref{pot}) with The main matrix of these systems is $
\left(\begin{array}{cc}
{\gamma}_{1} &{\gamma}_{2} \\
{\gamma}_{2} & {\gamma}_{1}
\end{array}\right)
$. By a linear change of variables, the matrix (\ref{pot}) can be reduced to following canonical Jordan form $\left(\begin{array}{cc}
{\gamma}_{1} + {\gamma}_{2}& 0 \\
0 &{\gamma}_{1} - {\gamma}_{2}
\end{array}\right)
$ Because of properties of symmetric systems, we will restrict our attention to $ \gamma_2\leq \gamma_1,~\gamma _{1,2}=0,1$. Similarly we will deal with two canonical Jordan form $\left(\begin{array}{cc}
1 &0 \\
0 & 1
\end{array}\right) $ and $ \left(\begin{array}{cc}
1 &0 \\
0 & 0
\end{array}\right)$. Imposing compatibility condition among the classes of systems and an arbitrary seventh order 0-homogeneous, we obtain a system of equations among the undetermined constants. If we separate out the coefficients of powers of $u$ and $v$ in this equation then in some condition the coefficients of $u_{nx}^m*v_{n'x}^{m'}$ all vanish identically. Solutions of the compatibility condition are given in the following theorems.

%%%%%%%%%%%%%%%%%%%%%%%%%%%%%%%%%%%%%%%%%%%%%%%%%%%%%%%%%%%%%%%%%%%%%%%%%%%%%%%%%%%%%%%%%%%%%%%%%%%%%%%%%%%%%%%%%%%%%%%%%%%%%%%%%%%%%%%%%%%%%%%%%%%%%%

\begin{theo}

A coupled fifth-order system of two-component evolution equations of the forms (\ref{pot}) and (\ref{five}) that possesses a seventh-order generalized symmetry of form (\ref{pot})   with $\gamma_1=\gamma_2=1$ have a lower order symmetry or can be  transformed by a linear change of variables to one of the following two systems (\ref{s1}) and (\ref{s2}).
\end{theo}

\begin{theo}
Every coupled fifth-order system of two-component evolution equations of form (\ref{pot}) and (\ref{five}) that possesses a seventh-order generalized symmetry of form (\ref{pot})   with $\gamma_1=1,~~\gamma_2=0$ have a lower order symmetry.
\end{theo}

\subsection {Integrability of the system (\ref{es1}) }$\\ $

System (\ref{es1})  possesses a symplectic operator as
\begin{eqnarray} 
S = \left(\begin{array}{cc} 
2D_x&D_x \\
 D_x& 2D_x
\end{array}\right).
\end{eqnarray}
Second Hamiltonian or symplectic operator for  this system is an open question for us.

\subsection {Integrability of the system (\ref{es2}) }$\\ $

Our main concern now is to show the integrability of the system (2.2).  To achieve this goal,  we set
\begin{eqnarray} 
 R = \left(\begin{array}{cc} 
R _{1}&R _{2}\\
R_{3}&R_4
\end{array}\right) 
\end{eqnarray}

 where
$\\
R_1= \alpha_{1} D_x^6 + \alpha_{2} D_x^4 + \alpha_{3} D_x^3 + \alpha_{4} D_x^2 + \alpha_{5} D_x + \alpha_{6} + \alpha_{01} D_x^{-1} \alpha_{07} + \alpha_{02} D_x^{-1} \alpha_{05} \\ \\
R_2= \alpha_{7} D_x^5 + \alpha_{8} D_x^4 + \alpha_{9} D_x^3 + \alpha_{10} D_x^2 + \alpha_{11} D_x + \alpha_{12} + \alpha_{01} D_x^{-1} \alpha_{08} + \alpha_{02} D_x^{-1} \alpha_{06} \\ \\
R_3= \alpha_{13} D_x^5 + \alpha_{14} D_x^4 + \alpha_{15} D_x^3 + \alpha_{16} D_x^2 + \alpha_{17} D_x + \alpha_{18}+ \alpha_{03} D_x^{-1} \alpha_{07} + \alpha_{04} D_x^{-1} \alpha_{05}\\ \\
R_4= \alpha_{19} D_x^4 + \alpha_{20} D_x^3 + \alpha_{21} D_x^2 + \alpha_{22} D_x + \alpha_{23} + \alpha_{03} D_x^{-1} \alpha_{08} + \alpha_{04} D_x^{-1} \alpha_{06} \\ \\
\alpha_{1}=12 \\\\
\alpha_{2}=72{u}_{1}-72u^2-30v^2 \\\\
\alpha_{3}=180{u}_{2}-360{u}_{1}u-18uv^2-60{v}_{1}v \\\\
\alpha_{4}=168{u}_{3}-480{u}_{2}u-372{u}_{1}^2-72{u}_{1}u^2-126{u}_{1}v^2+108u^4+114u^2v^2-48u{v}_{1}v-72{v}_{2}v-72{v}_{1}^2+12v^4 \\\\
\alpha_{5}=72{u}_{4}-360{u}_{3}u-756{u}_{2}{u}_{1}-108{u}_{2}u^2-126{u}_{2}v^2-216{u}_{1}^2u+648{u}_{1}u^3+342{u}_{1}uv^2-180{u}_{1}{v}_{1}v+18u^3v^2+180u^2{v}_{1}v-72u{v}_{2}v-72u{v}_{1}^2+9uv^4-48{v}_{3}v-144{v}_{2}{v}_{1}+54{v}_{1}v^3 \\\\
\alpha_{6}=12{u}_{5}-144{u}_{4}u-276{u}_{3}{u}_{1}-36{u}_{3}u^2-36{u}_{3}v^2-180{u}_{2}^2-456{u}_{2}{u}_{1}u+456{u}_{2}u^3+210{u}_{2}uv^2-96{u}_{2}{v}_{1}v-72{u}_{1}^3+888{u}_{1}^2u^2+132{u}_{1}^2v^2+108{u}_{1}u^2v^2+288{u}_{1}u{v}_{1}v-84{u}_{1}{v}_{2}v-84{u}_{1}{v}_{1}^2+18{u}_{1}v^4-48u^6-84u^4v^2+48u^3{v}_{1}v+156u^2{v}_{2}v+156u^2{v}_{1}^2-30u^2v^4-72u{v}_{3}v-216u{v}_{2}{v}_{1}+78u{v}_{1}v^3-12{v}_{4}v-48{v}_{3}{v}_{1}-36{v}_{2}^2+24{v}_{2}v^3+36{v}_{1}^2v^2 \\\\
\alpha_{7}=12v \\\\
\alpha_{8}=-24uv+60{v}_{1} \\\\
\alpha_{9}=-48{u}_{1}v-24u^2v-96u{v}_{1}+120{v}_{2}-30v^3 \\\\
\alpha_{10}=-60{u}_{2}v-72{u}_{1}uv-144{u}_{1}{v}_{1}+48u^3v-72u^2{v}_{1}-144u{v}_{2}+42uv^3+120{v}_{3}-150{v}_{1}v^2 \\\\
 \alpha_{11}=-72{u}_{3}v-72{u}_{2}uv-120{u}_{2}{v}_{1}-84{u}_{1}^2v+216{u}_{1}u^2v-144{u}_{1}u{v}_{1}-144{u}_{1}{v}_{2}+54{u}_{1}v^3+12u^4v+96u^3{v}_{1}-72u^2{v}_{2}+30u^2v^3-96u{v}_{3}+156u{v}_{1}v^2+60{v}_{4}-162{v}_{2}v^2-192{v}_{1}^2v+12v^5 \\\\
 \alpha_{12}=-48{u}_{4}v-24{u}_{3}uv-72{u}_{3}{v}_{1}-228{u}_{2}{u}_{1}v+180{u}_{2}u^2v-72{u}_{2}u{v}_{1}-60{u}_{2}{v}_{2}+54{u}_{2}v^3+240{u}_{1}^2uv-84{u}_{1}^2{v}_{1}+24{u}_{1}u^3v+216{u}_{1}u^2{v}_{1}-72{u}_{1}u{v}_{2}+66{u}_{1}uv^3-48{u}_{1}{v}_{3}+114{u}_{1}{v}_{1}v^2-24u^5v+12u^4{v}_{1}+48u^3{v}_{2}-42u^3v^3-24u^2{v}_{3}+66u^2{v}_{1}v^2-24u{v}_{4}+114u{v}_{2}v^2+144u{v}_{1}^2v-15uv^5+12{v}_{5}-78{v}_{3}v^2-276{v}_{2}{v}_{1}v-72{v}_{1}^3+66{v}_{1}v^4 \\\\
\alpha_{13}=-6v \\\\
\alpha_{14}=6uv+12{v}_{1} \\\\
\alpha_{15}=-36{u}_{1}v+30u^2v-12u{v}_{1}+15v^3 \\\\
\alpha_{16}=-54{u}_{2}v+150{u}_{1}uv+84{u}_{1}{v}_{1}-30u^3v-60u^2{v}_{1}+21uv^3+24{v}_{1}v^2 \\\\
\alpha_{17}=-30{u}_{3}v+144{u}_{2}uv+24{u}_{2}{v}_{1}+78{u}_{1}^2v-54{u}_{1}u^2v-180{u}_{1}u{v}_{1}+45{u}_{1}v^3-24u^4v+60u^3{v}_{1}-24u^2v^3+30u{v}_{1}v^2+18{v}_{2}v^2-30{v}_{1}^2v-6v^5 \\\\
\alpha_{18}=-6{u}_{4}v+66{u}_{3}uv+36{u}_{3}{v}_{1}+90{u}_{2}{u}_{1}v-36{u}_{2}u^2v-108{u}_{2}u{v}_{1}+18{u}_{2}v^3+48{u}_{1}^2uv+24{u}_{1}^2{v}_{1}-144{u}_{1}u^3v-72{u}_{1}u^2{v}_{1}-72{u}_{1}uv^3+18{u}_{1}{v}_{1}v^2+24u^5v+48u^4{v}_{1}-12u^3v^3-48u^2{v}_{1}v^2+18u{v}_{2}v^2-42u{v}_{1}^2v-12uv^5+6{v}_{3}v^2+30{v}_{2}{v}_{1}v+12{v}_{1}^3-18{v}_{1}v^4 \\\\
\alpha_{19}=-6v^2 \\\\
\alpha_{20}=18uv^2-12{v}_{1}v \\\\
\alpha_{21}=12{u}_{1}v^2-6u^2v^2+18u{v}_{1}v-36{v}_{2}v+36{v}_{1}^2+15v^4 \\\\
\alpha_{22}=18{u}_{2}v^2+18{u}_{1}uv^2+36{u}_{1}{v}_{1}v-18u^3v^2+54u{v}_{2}v-72u{v}_{1}^2-9uv^4-24{v}_{3}v+36{v}_{2}{v}_{1}+54{v}_{1}v^3 \\\\
\alpha_{23}=18{u}_{3}v^2-30{u}_{2}{v}_{1}v+30{u}_{1}^2v^2-66{u}_{1}u^2v^2-42{u}_{1}u{v}_{1}v+12{u}_{1}{v}_{2}v+12{u}_{1}{v}_{1}^2-15{u}_{1}v^4+12u^4v^2+18u^3{v}_{1}v-6u^2{v}_{2}v+12u^2{v}_{1}^2-6u^2v^4+18u{v}_{3}v-36u{v}_{2}{v}_{1}-27u{v}_{1}v^3-6{v}_{4}v+12{v}_{3}{v}_{1}+33{v}_{2}v^3-6{v}_{1}^2v^2-6v^6 \\\\
\alpha_{01}=-24{u}_{5}-120{u}_{3}{u}_{1}+120{u}_{3}u^2+48{u}_{3}v^2-120{u}_{2}^2+480{u}_{2}{u}_{1}u+36{u}_{2}uv^2+48{u}_{2}{v}_{1}v+120{u}_{1}^3+72{u}_{1}^2v^2-120{u}_{1}u^4-144{u}_{1}u^2v^2+48{u}_{1}u{v}_{1}v+72{u}_{1}{v}_{2}v+72{u}_{1}{v}_{1}^2-6{u}_{1}v^4-48u^3{v}_{1}v+24u^2{v}_{2}v+24u^2{v}_{1}^2+48u{v}_{3}v+144u{v}_{2}{v}_{1}-60u{v}_{1}v^3-24{v}_{4}v-96{v}_{3}{v}_{1}-72{v}_{2}^2+48{v}_{2}v^3+72{v}_{1}^2v^2 \\\\
\alpha_{02}=-3{u}_{1} \\\\
\alpha_{03}=12{u}_{4}v-12{u}_{3}uv-24{u}_{3}{v}_{1}+60{u}_{2}{u}_{1}v-48{u}_{2}u^2v+24{u}_{2}u{v}_{1}-24{u}_{2}v^3-96{u}_{1}^2uv-72{u}_{1}^2{v}_{1}+48{u}_{1}u^3v+96{u}_{1}u^2{v}_{1}-48{u}_{1}uv^3-36{u}_{1}{v}_{1}v^2-24u^4{v}_{1}-36u{v}_{2}v^2+36u{v}_{1}^2v+12{v}_{3}v^2+12{v}_{2}{v}_{1}v-24{v}_{1}^3-30{v}_{1}v^4 \\\\
\alpha_{04}=-3{v}_{1} \\\\
\alpha_{05}=8u_4+8v_3v+40u_2u_1-40u_2u^2-4u_2v^2+24v_2v_1-16v_2uv-40u_1^2u-8u_1v_1v-16v_1^2u-8v_1u^2v-4v_1v^3+8u^5+8u^3v^2+2uv^4 \\\\
\alpha_{06}=-8u_3v-16u_2uv-8v_2v^2-12u_1^2v+8u_1u^2v+4u_1v^3-8v_1^2v+4u^4v+4u^2v^3+v^5 \\\\
\alpha_{07}=u \\\\
\alpha_{08}=\frac{v}{2} \\ \\$

This shows that the system (\ref{es2}) passes the integrability test.

\subsection {Integrability of system (\ref{s1}) }$\\ $

We proceed as before to  show the integrability of the system (\ref{s1}).  By change of dependent variables
\begin{eqnarray}
u \rightarrow \frac{1}{2}\int(w-z)dx\\
v\rightarrow \frac{1}{2}\int(w+z)dx
\end{eqnarray}
system (\ref{s1}) can be written in its canonical form as
\begin{eqnarray}
\left(
\begin{array}{l}
u\\ \\\\\\\\
v\\ \\\\
\end{array}\right)_{t}=
\left(
\begin{array}{cc}
w_{5x}-2zz_{4x}-10w_{x}w_{3x}-20w^2w_{3x}-2z^2w_{3x}-8z_{x}z_{3x}\\-8wzz_{3x} -10w_{xx}^2 -80ww_{x}w_{xx}-8zz_{x}w_{xx} -6z_{xx}^2\\-12zw_{x}z_{xx} -24wz_{x}z_{xx} +8w^2zz_{xx} +4z^3z_{xx}-20w_{x}^3-12w_{x}z_{x}^2\\ +16wzw_{x}z_{x}+80w^4w_{x} +48w^2z^2w_{x} +4z^4w_{x}+8w^2z_{x}^2\\+12z^2z_{x}^2 +32w^3zz_{x}+16wz^3z_{x} \\\\
4zw_{4x}+4z_{x}w_{3x}-16wzw_{3x}-8z^2z_{3x}-40zw_{x}w_{xx}\\-16wz_{x}w_{xx} -16w^2zw_{xx} -8z^3w_{xx} -32zz_{x}z_{xx} -12w_{x}^2z_{x} \\-32wzw_{x}^2 -16w^2w_{x}z_{x} -24z^2w_{x}z_{x} +64w^3zw_{x}+32wz^3w_{x}\\-8z_{x}^3+16w^4z_{x} +48w^2z^2z_{x} +20z^4z_{x}
\end{array}\right)\label{js1}
\end{eqnarray}
{\bf Proposition} The infinite hierarchy of the system (\ref{js1}) can be written in two different ways
\begin{eqnarray}
\left(
\begin{array}{cc}
w_{t}\\
z_{t}
\end{array}
\right)
=
\mathrm{J}
\left(
\begin{array}{cc}
\delta_{w}\\
\delta_{z}
\end{array}
\right)
\int \rho_{1}~ \mathrm{d}x
=\mathrm{K}
\left(
\begin{array}{cc}
\delta_{w}\\
\delta_{z}
\end{array}
\right)
\int \rho_{0} ~\mathrm{d}x
\end{eqnarray}
with the compatible pair of Hamiltonian operators
\begin{eqnarray}
\mathrm{J}= \left(
\begin{array}{cc}
D_x &0 \\
0& 2D_x
\end{array}\right),\mathrm{K} = \left(
\begin{array}{cc}
\mathrm{K} _{1}^1&\mathrm{K} _{2}^1\\
\mathrm{K}_{3}^1&\mathrm{K} _{4}^1
\end{array}\right)
\end{eqnarray}
where
\begin{eqnarray}
\begin{array}{ll}
K_1^1=&D_x^7+\omega_1D_x^5+D_x^5\omega_1+\omega_2D_x^3+D_x^3\omega_2+\omega_3D_x+D_x\omega_3+8w_{x}D_x^{-1}w_t\\&+8w_tD_x^{-1}w_{x}
\\
K_2^1=&D_x^6\omega_4+D_x^5\omega_5+D_x^4\omega_6+D_x^3\omega_7+D_x^2\omega_8+D_x\omega_9+\omega_{10}+8w_{x}D_x^{-1}z_t\\&+8w_tD_x^{-1}z_{x} \\
K_3^1=& -\omega_4D_x^6+ \omega_5D_x^5- \omega_6D_x^4+ \omega_7D_x^3- \omega_8D_x^2+ \omega_9D_x-\omega_{10} +8z_{x}D_x^{-1}w_t\\&+8z_tD_x^{-1}w_{x} \\
k_4^1=&\omega_{11}D_x^5+D_x^5\omega_{11}+\omega_{12}D_x^3+D_x^3\omega_{12}+\omega_{13}D_x+D_x\omega_{13} +8z_{x}D_x^{-1}z_t\\&+8z_tD_x^{-1}z_{x} \\
\end{array}
\end{eqnarray}
and the coefficients satisfy
\begin{eqnarray}
\begin{array}{ll}
\omega_1=&-6w_{x}-12w^2-2z^2 \\
\omega_2=&16w_{3x}+40ww_{xx}+8zz_{xx}+58w_{x}^2+24w^2w_{x}+12z^2w_{x}+8z_{x}^2+16wzz_{x}\\&+72w^4 +40w^2z^2 +18z^4 \\
\omega_3=&-10w_{5x}-24ww_{4x}-4zz_{4x}-100w_{x}w_{3x}-24w^2w_{3x}-12z^2w_{3x}-16z_{x}z_{3x}\\&-84w_{xx}^2-64ww_{x}w_{xx} -64zz_{x}w_{xx}-128w^3w_{xx}-64wz^2w_{xx}-12z_{xx}^2\\&-56zw_{x}z_{xx}-16w^2zz_{xx}-152z^3z_{xx} -48w_{x}^3 -704w^2w_{x}^2-80z^2w_{x}^2\\&-56w_{x}z_{x}^2-288wzw_{x}z_{x} -96z^4w_{x} -16w^2z_{x}^2-216z^2z_{x}^2\\& -64w^3zz_{x} +96wz^3z_{x} -128w^6-128w^4z^2-96w^2z^4-16z^6\\
\omega_4=&-4z \\
\omega_5=&4z_{x}-16wz \\
\omega_6=&+48zw_{x}+16wz_{x}+32w^2z \\
\omega_7=&-72zw_{xx}-32w_{x}z_{x}-160wzw_{x}-32w^2z_{x}+96z^2z_{x}+128w^3z \\
\omega_8=&+40zw_{3x}+40z_{x}w_{xx}+192wzw_{xx}+208zw_{x}^2+96ww_{x}z_{x}-576w^2zw_{x}\\&-320zz_{x}^2 -128w^3z_{x} -64w^4z \\
\omega_9=&-8zw_{4x}-128wzw_{3x}-104z^2z_{3x}-240zw_{x}w_{xx}-96wz_{x}w_{xx}+480w^2zw_{xx}\\&-96z^3w_{xx} -152zz_{x}z_{xx}+576wz^2z_{xx}-112w_{x}^2z_{x}+640wzw_{x}^2+192w^2w_{x}z_{x}\\&-192z^2w_{x}z_{x} +384w^3zw_{x} +384wz^3w_{x}+160z_{x}^3+448wzz_{x}^2+64w^4z_{x}\\&-768w^2z^2z_{x}-96z^4z_{x} -256w^5z \\
\omega_{10}=&+8z_{x}w_{4x}+32wzw_{4x}+40z^2z_{4x}+32wz_{x}w_{3x}-128w^2zw_{3x}+48z^3w_{3x}\\&+224zz_{x}z_{3x} -288wz^2z_{3x} -80w_{x}z_{x}w_{xx}-320wzw_{x}w_{xx}-288w^2z_{x}w_{xx}\\&-128w^3zw_{xx}-192wz^3w_{xx} +304z^2z_{x}w_{xx} +120zz_{xx}^2 +64z^2w_{x}z_{xx}\\&+192z_{x}^2z_{xx} -1376wzz_{x}z_{xx}+384w^2z^2z_{xx} +48z^4z_{xx} -256ww_{x}^2z_{x}\\& -256w^2zw_{x}^2 +64z^3w_{x}^2 +160zw_{x}z_{x}^2 -128w^3w_{x}z_{x} -192wz^2w_{x}z_{x}\\& +512w^4zw_{x}-512wz_{x}^3 +1216w^2zz_{x}^2+272z^3z_{x}^2+256w^5z_{x} \\
\omega_{11}=&-12z^2 \\
\omega_{12}=&+100zz_{xx}-72z^2w_{x}+80z_{x}^2-96wzz_{x}+144w^2z^2+36z^4 \\
\omega_{13}=&-68zz_{4x}+8z^2w_{3x}-292z_{x}z_{3x}+176wzz_{3x}+232zz_{x}w_{xx}+64wz^2w_{xx}\\&-264z_{xx}^2 +512zw_{x}z_{xx} +816wz_{x}z_{xx} -896w^2zz_{xx}-88z^3z_{xx}-64z^2w_{x}^2\\&+280w_{x}z_{x}^2 -1056wzw_{x}z_{x} +96z^4w_{x} -752w^2z_{x}^2 -380z^2z_{x}^2\\&+768w^3zz_{x}+96wz^3z_{x}-384w^4z^2 -192w^2z^4-32z^6\\
\end{array}
\end{eqnarray}

The first few conserved densities of the hierarchy are listed as follows
\begin{eqnarray}
\begin{array}{ll}
\rho_{0}=&\alpha\\
\rho_1=&2w^2+z^2\\
\rho_2=&+3ww_{4x}+10w^2w_{3x}+6z^2w_{3x}-20w^3w_{xx}-6wz^2w_{xx}-18w^2zz_{xx}-4z^3z_{xx}\\&-18w^2z_{x}^2 +16w^3zz_{x} +24wz^3z_{x}+16w^6+24w^4z^2+12w^2z^4+2z^6\\
 
\dot{.}&\\
\dot{.}&\\
\end{array}
\end{eqnarray}
These densities are sufficient to write two Magri schemes with the same Hamiltonian operators such that one of them contains the new system, and this confirms the integrability of the system (\ref{js2}).

\subsection {Integrability of system (\ref{s2}) }$\\ $

In a manner parallel to the analysis presented earlier, and to prove the integrability of the system (\ref{s2}), we use the  change of dependent variables
\begin{eqnarray}
u \rightarrow \frac{1}{2}\int(w-z)dx\\
v\rightarrow \frac{1}{2}\int(w+z)dx
\end{eqnarray}
which carries the system (\ref{s2}) to its canonical form as
\begin{eqnarray}
\left(
\begin{array}{l}
u\\ \\\\\\\\
v\\ \\\\
\end{array}\right)_{t}=
\left(
\begin{array}{cc}
w_{5x}-zz_{4x}+10w_{x}w_{3x}-20w^2w_{3x}-8z^2w_{3x}-4z_{x}z_{3x}-2wzz_{3x}\\+10w_{xx}^2 -80ww_{x}w_{xx} -32zz_{x}w_{xx}-3z_{xx}^2 -18zw_{x}z_{xx}-6wz_{x}z_{xx} \\+16w^2zz_{xx} +8z^3z_{xx} -20w_{x}^3-18w_{x}z_{x}^2 +32wzw_{x}z_{x}+80w^4w_{x}\\+48w^2z^2w_{x} +4z^4w_{x}+16w^2z_{x}^2+24z^2z_{x}^2 +32w^3zz_{x}+16wz^3z_{x}\\ \\
2zw_{4x}+2z_{x}w_{3x}-4wzw_{3x}-2z^2z_{3x}+20zw_{x}w_{xx}-4wz_{x}w_{xx} \\-32w^2zw_{xx} -16z^3w_{xx} -8zz_{x}z_{xx}+12w_{x}^2z_{x} -64wzw_{x}^2\\ -48z^2w_{x}z_{x} +64w^3zw_{x} -32w^2w_{x}z_{x}+32wz^3w_{x}-2z_{x}^3\\+16w^4z_{x} +48w^2z^2z_{x} +20z^4z_{x}
\end{array}\right)\label{js2}
\end{eqnarray}
{\bf Proposition} The infinite hierarchy of system (\ref{js2}) can be write in not just one but two different ways
\begin{eqnarray}
\left(
\begin{array}{cc}
w_{t}\\
z_{t}
\end{array}
\right)
=
\mathrm{J}
\left(
\begin{array}{cc}
\delta_{w}\\
\delta_{z}
\end{array}
\right)
\int \rho_{1}~ \mathrm{d}x
=\mathrm{K}
\left(
\begin{array}{cc}
\delta_{w}\\
\delta_{z}
\end{array}
\right)
\int \rho_{1} ~\mathrm{d}x
\end{eqnarray}
with the compatible pair of Hamiltonian operators
\begin{eqnarray}
\mathrm{J}= \left(
\begin{array}{cc}
D_x &0 \\
0& 2D_x
\end{array}\right),\mathrm{K^2} = \left(
\begin{array}{cc}
\mathrm{K}_{1}^2&\mathrm{K}_{2}^2\\
\mathrm{K}_{3}^2&\mathrm{K}_{4}^2
\end{array}\right)
\end{eqnarray}
where
\begin{eqnarray}
\begin{array}{ll}
K_1^2=&D_x^7+\psi_1D_x^5+D_x^5\psi_1+\psi_2D_x^3+D_x^3\psi_2+\psi_3D_x+D_x\psi_3+8w_xD_x^{-1}w_t\\&+8w_tD_x^{-1}w_x\\ 
K_2^2=&D_x^6\psi_4+D_x^5\psi_5+D_x^4\psi_6+D_x^3\psi_7+D_x^2\psi_8+D_x\psi_9+\psi_{10}+8w_xD_x^{-1}z_t\\&+8w_tD_x^{-1}z_x\\ 
K_3^2=&-\psi_4D_x^6+\psi_5D_x^5-\psi_6D_x^4+\psi_7D_x^3-\psi_8D_x^2+\psi_9D_x-\psi_{10}+8z_xD_x^{-1}w_t\\&+8z_tD_x^{-1}w_x\\ 
K_4^2=&\psi_{11}D_x^5+D_x^5\psi_5+\psi_{12}D_x^3+D_x^3\psi_{12}+\psi_{13}D_x+D_x\psi_{13}+8z_xD_x^{-1}z_t\\&+8z_tD_x^{-1}z_x 
\end{array}
\end{eqnarray}
where the coefficients satisfy
\begin{eqnarray}
\begin{array}{ll}
\psi_1=&6w_{x}-12w^2-5z^2\\
\psi_2=&-16w_{3x}+40ww_{xx}+26zz_{xx}+58w_{x}^2-24w^2w_{x}-12z^2w_{x}+26z_{x}^2\\&+20wzz_{x} +72w^4+52w^2z^2+18z^4\\
\end{array}
\end{eqnarray}
\begin{eqnarray}
\begin{array}{ll}
\psi_3=&10w_{5x}-24ww_{4x}-16zz_{4x}-100w_{x}w_{3x}+24w^2w_{3x}+12z^2w_{3x}-64z_{x}z_{3x}\\&-12wzz_{3x} -84w_{xx}^2+64ww_{x}w_{xx}+28zz_{x}w_{xx}-128w^3w_{xx}-40wz^2w_{xx}\\&-48z_{xx}^2-16zw_{x}z_{xx} -36wz_{x}z_{xx}-88w^2zz_{xx}-134z^3z_{xx} +48w_{x}^3\\&-704w^2w_{x}^2-104z^2w_{x}^2-16w_{x}z_{x}^2 -288wzw_{x}z_{x} +24z^4w_{x}-88w^2z_{x}^2\\&-216z^2z_{x}^2 -128w^3zz_{x}-24wz^3z_{x}-128w^6 -128w^4z^2-96w^2z^4-16z^6\\
\psi_4=&-2z\\
\psi_5=&2z_{x}-4wz\\
\psi_6=&-24zw_{x}+4wz_{x}+40w^2z\\
\psi_7=&+36zw_{xx}+28w_{x}z_{x}-200wzw_{x}-40w^2z_{x}+120z^2z_{x}+80w^3z\\
\psi_8=&-20zw_{3x}-8z_{x}w_{xx}+192wzw_{xx}+104zw_{x}^2+120ww_{x}z_{x}-144w^2zw_{x}\\&-80w^3z_{x}-424zz_{x}^2-128w^4z\\

\psi_9=&+4zw_{4x}+12z_{x}w_{3x}-88wzw_{3x}-154z^2z_{3x}-120zw_{x}w_{xx}-72wz_{x}w_{xx}\\&+96w^2zw_{xx} +48z^3w_{xx} -238zz_{x}z_{xx}+360wz^2z_{xx}+16w_{x}^2z_{x}-128wzw_{x}^2\\&-96w^2w_{x}z_{x} +312z^2w_{x}z_{x}+768w^3zw_{x} -96wz^3w_{x}+224z_{x}^3 +232wzz_{x}^2\\&+128w^4z_{x} -672w^2z^2z_{x}-192z^4z_{x}-256w^5z\\

\psi_{10}=&+8z_{x}w_{4x}+16wzw_{4x}+62z^2z_{4x}+16wz_{x}w_{3x}-32w^2zw_{3x}-24z^3w_{3x}\\&-180wz^2z_{3x} +364zz_{x}z_{3x} +80w_{x}z_{x}w_{xx}+160wzw_{x}w_{xx}-296z^2z_{x}w_{xx}\\&-256w^3zw_{xx}-192w^2z_{x}w_{xx} +48wz^3w_{xx}+186zz_{xx}^2 -176z^2w_{x}z_{xx}\\&+348z_{x}^2z_{xx}-812wzz_{x}z_{xx}+336w^2z^2z_{xx} +96z^4z_{xx}-64ww_{x}^2z_{x}\\&-512w^2zw_{x}^2 +128z^3w_{x}^2 -656zw_{x}z_{x}^2-256w^3w_{x}z_{x} +512w^4zw_{x}\\&+912wz^2w_{x}z_{x}-272wz_{x}^3+1136w^2zz_{x}^2+544z^3z_{x}^2+256w^5z_{x}\\

\psi_{11}=&-12z^2\\
\psi_{12}=&+190zz_{xx}-144z^2w_{x}+74z_{x}^2-120wzz_{x}+144w^2z^2+36z^4\\
\psi_{13}=&-146zz_{4x}+88z^2w_{3x}-514z_{x}z_{3x}+244wzz_{3x}+560zz_{x}w_{xx}-224wz^2w_{xx}\\&-480z_{xx}^2 +904zw_{x}z_{xx} +1092wz_{x}z_{xx}-824w^2zz_{xx}-160z^3z_{xx}+512w_{x}z_{x}^2\\&-520z^2w_{x}^2 -1632wzw_{x}z_{x}+864w^2z^2w_{x}+192z^4w_{x} -632w^2z_{x}^2-464z^2z_{x}^2\\&+672w^3zz_{x} +192wz^3z_{x}-384w^4z^2-192w^2z^4-32z^6\\
\end{array}
\end{eqnarray}

The first few conserved densities of the system (\ref{js2}) are listed as follows
\begin{eqnarray}
\begin{array}{ll}
\rho_{0}=&\alpha\\
\rho_1=&2w^2+z^2\\
\rho_2=&+3ww_{4x}-10w^2w_{3x}+3z^2w_{3x}-20w^3w_{xx}-24wz^2w_{xx}+18w^2zz_{xx}\\&-z^3z_{xx} +18w^2z_{x}^2 +32w^3zz_{x}+48wz^3z_{x}+16w^6+24w^4z^2+12w^2z^4+2z^6\\
\dot{.}&\\
\dot{.}&\\
\end{array}
\end{eqnarray}
These densities suffice to write two Magri schemes with same Hamiltonian operators that one of them contains the new system, and this in turn emphasizes  the integrability of the system (\ref{js2}).

\section{Discussion}

In this work we established four    fifth-order integrable coupled systems of weight $0$ and $1$. We examined the  related recursion operator and bi-Hamiltonian formulations for the developed systems.   We used the compatible pair of Hamiltonian operators to formally prove  the integrability of the developed systems.  The obtained results will add valuable findings to the existing integrable systems of fifth order two-component equation.  It is expected that other works will be conducted for  recovering the scientific features of these systems of   equations.

\end{document}